\documentstyle[psfig]{czjphys}         % for LaTeX 2.09
\begin{document}
\title{Generation of Bianchi type V cosmological models 
with varying $\Lambda$-term }
%\footnote{
%       the title, authors, and addresses, the footnotemarks
%       [$^*),\ ^{**}),\ ^{\dagger})$, etc.] are used 
%       automatically.}}
%
\authori{ANIRUDH PRADHAN, \footnote{E-mail:pradhan@iucaa.ernet.in, acpradhan@yahoo.com}}      
\addressi{Department of Mathematics, Hindu Post-graduate College , 
Zamania-232 331, Ghazipur, India}
\authorii{ANIL KUMAR YADAV, LALLAN YADAV}
\addressii{Department of Physics, K.N.Govt. Post-graduate College, 
Gyanpur, Sant Ravidas Nagar, Bhadohi - 221 304, India }
\authoriii{}    \addressiii{}
\authoriv{}     \addressiv{}
\authorv{}      \addressv{}
\authorvi{}     \addressvi{}
\headauthor{A. Pradhan et al.}   
\headtitle{Generation of Bianchi type V  \ldots}
\lastevenhead{A. Pradhan et al.: Generation of Bianchi type V 
              ldots}
%%%%%%%%%%%%%% FOR EDITORIAL USE ONLY!!! %%%%%%%%%%%%%%%
\refnum{A}%\total{}\type{}
\daterec{XXX}    %;\\ final version }
\issuenumber{0}  \year{2004}
\setcounter{page}{1}
%\firstpage{1}
%\lastpage{000}
%\makefirsttitle
%%%%%%%%%%%%%%%%%%%%%%%%%%%%%%%%%%%%%%%%%%%%%%%%%%%%%%%%
\maketitle
%Page headings:
\begin{abstract}
Bianchi type V perfect fluid cosmological models are investigated with 
cosmological term $\Lambda$ varying with time. Using a generation 
technique (Camci {\it et al.}, 2001), it is shown that the Einstein's field 
equations are solvable for any arbitrary cosmic scale function. Solutions for 
particular forms of cosmic scale functions are also obtained. The cosmological 
constant is found to be decreasing function of time, which is supported by 
results from recent type Ia supernovae observations. Some physical aspects of the 
models are also discussed.  
\end{abstract}

\section{Introduction}     %\section*{Introduction}

The study of Bianchi type V cosmological models create more interest as these 
models contain isotropic special cases and permit arbitrary small anisotropy 
levels at some instant of cosmic time. This property makes them suitable as
model of our universe. The homogeneous and isotropic Friedman-Robertson-Walker
(FRW) cosmological models, which are used to describe standard cosmological models,
are particular case of Bianchi type I, V and IX universes, according to whether 
the constant curvature of the physical three-space, $t$ = constant, is zero, 
negative or positive. These models will be interesting to construct cosmological
models of the types which are of class one. Present cosmology is based on the
FRW model which is completely homogeneous and isotropic. This is in agreement with 
observational data about the large scale structure of the universe. However, although 
homogeneous but anisotropic models are more restricted than the inhomogeneous models,
they explain a number of observed phenomena quite satisfactorily. This stimulates the 
research for obtaining exact anisotropic solution for Einstein's field equations 
(EFEs) as a cosmologically accepted physical models for the universe (at least in 
the early stages). Roy and Prasad [1] have investigated Bianchi type V 
universes which are locally rotationally symmetric and are of embedding class one 
filled with perfect fluid with heat conduction and radiation. Bianchi type V 
cosmological models have been studied by other researchers (Farnsworth [2], 
Maartens and Nel [3], Wainwright {\it et al.} [4], Collins [5], 
Meena and Bali [6], Pradhan {\it et al.} [7, 8]) in different context.    

Models with a relic cosmological constant $\Lambda$ have received considerable 
attention recently among researchers for various reasons 
(see Refs. [9 - 13] and references therein). Some of the 
recent discussions on the cosmological constant ``problem'' and on cosmology 
with a time-varying cosmological constant by Ratra and Peebles [14], 
Dolgov [15 - 17] and Sahni and Starobinsky [18]
point out that in the absence of any interaction with matter or radiation, the 
cosmological constant remains a ``constant'', however, in the presence of
interactions with matter or radiation, a solution of Einstein equations and the 
assumed equation of covariant conservation of stress-energy with a time-varying 
$\Lambda$ can be found. For these solutions, conservation of energy requires 
decrease in the energy density of the vacuum component to be compensated by a 
corresponding increase in the energy density of matter or radiation. Earlier 
researchers on this topic, are contained in Zeldovich [19], Weinberg [10] and Carroll,
Press and Turner [20]. Recent cosmological observations by High-z Supernova Team and 
Supernova Cosmological Project (Garnavich {\it et al.} [21], Perlmutter {\it et al.} [22], 
Riess {\it et al.} [23], Schmidt {\it et al.} [24])  strongly favour a significant and 
positive $\Lambda$ with the magnitude $\Lambda(G\hbar / c^{3}) \approx 10^{-123}$. Their 
finding arise from the study of more than $50$ type Ia supernovae with redshifts in 
the range $0.10 \leq z \leq 0.83$ and suggest Friedman models with negative pressure
matter such as a cosmological constant, domain walls or cosmic strings (Vilenkin [25],
Garnavich {\it et al.} [21]. The main conclusion of these observations on magnitude and 
red-shift of type Ia supernova suggest that the expansion of the universe may be an 
accelerating one with a large function of cosmological density in the form of the 
cosmological $\Lambda$-term.

Several ans$\ddot{a}$tz have been proposed in which the $\Lambda$ term decays 
with time (see Refs. Gasperini [26], Freese {\it et al.} [27], $\ddot{O}$zer and Taha [13], 
Peebles and Ratra [28], Chen and Hu [29], Abdussattar and Viswakarma [30], Gariel and 
Le Denmat [31], Pradhan {\it et al.} [32]). Of the special interest is the ansatz 
$\Lambda \propto S^{-2}$ (where $S$ is the scale factor of the Robertson-Walker metric) 
by Chen and Wu [29], which has been considered/modified by several authors 
( Abdel-Rahaman [33], Carvalho {\it et al.} [13], Waga [34], Silveira and Waga [35],
Vishwakarma [36]). 

In recent years, several authors (Hajj-Boutros [37], Hajj-Boutros and
Sfeila [38], Ram [39], Mazumder [40] and Pradhan and
Kumar [41]) have investigated the solutions of EFEs for homogeneous but 
anisotropic models by using some different generation techniques. Bianchi spaces 
$I-IX$ are useful tools in constructing models of spatially homogeneous cosmologies
(Ellis and MacCallum [42], Ryan and Shepley [43]). From these
models, homogeneous Bianchi type V universes are the natural generalization of the 
open FRW model which eventually isotropize. Recently Camci {\it et al.} [44]
derived a new technique for generating exact solutions of EFEs with perfect fluid 
for Bianchi type V spacetime. In this paper, in what follows, we will discuss Bianchi 
type V cosmological models obtained by augmenting the energy-momentum tensor of a 
perfect fluid by a term that represents a variable cosmological constant varying with 
time, and later generalize the solutions of Refs. [39, 44]. This paper is 
organized as follows: The field equations and the generation technique are presented
in Section $2$. We relate three of the metric variables by solving the off-diagonal 
component of EFEs, and find a second integral which is used to relate the remaining
two metric variables. In Section 3, for the particular form of each metric variables, 
some solutions are presented separately and solutions of Camci {\it et al.} [44]
and Ram [39] are shown to be particular cases of our solutions. Kinematical
and dynamical properties of all solutions are also studied in this section. In Section 
$4$, we give the concluding remarks. 

%%%%%%%%%%%%%%%%%%%%%%%%%%%%%%%%%%%%%% SECTION 2 %%%%%%%%%%%%%%%%%%%%%%%%%%%%%%%%%%%%%%%
\section{Field equations and generation technique} 
In this section, we review the solutions obtained by Camci {\it et al.} [44].
The usual energy-momentum tensor is modified by addition of a term
\begin{equation}
\label{eq1}
T^{(vac)}_{ij} = - \Lambda(t) g_{ij},
\end{equation}
where $\Lambda(t)$ is the cosmological term and $g_{ij}$ is the metric tensor.
Thus the new energy-momentum tensor is
\begin{equation}
\label{eq2}
T_{ij} = (p + \rho)u_{i}u_{j} - p g_{ij} - \Lambda(t) g_{ij},
\end{equation}
where $\rho$ and $p$, respectively, the energy and pressure of cosmic fluid, and 
$u_{i}$ is the fluid four-velocity such that $u^{i} u_{i} = 1$. \\
We consider the spacetime metric of the spatially homogeneous Bianchi type V of
the form
\begin{equation}
\label{eq3}
ds^{2} = dt^{2} - A^{2}(t)dx^{2} - e^{2\alpha x}\left[B^{2}(t) dy^{2} + C^{2}(t) 
dz^{2}\right],
\end{equation}
where $\alpha$ is a constant. For the energy momentum tensor (\ref{eq2}) and Bianchi 
type V spacetime (\ref{eq3}),
Einstein's field equations
\begin{equation}
\label{eq4}
R_{ij} - \frac{1}{2} R g_{ij} = - 8\pi T_{ij}
\end{equation}
yield the following five independent equations
\begin{equation}
\label{eq5}
\frac{A_{44}}{A} + \frac{B_{44}}{B} + \frac{A_{4}B_{4}}{AB} - \frac{\alpha^{2}}
{A^{2}} = - 8\pi (p + \Lambda),
\end{equation}
\begin{equation}
\label{eq6}
\frac{A_{44}}{A} + \frac{C_{44}}{C} + \frac{A_{4}C_{4}}{AC} - \frac{\alpha^{2}}
{A^{2}} = - 8\pi (p + \Lambda),
\end{equation}
\begin{equation}
\label{eq7}
\frac{B_{44}}{B} + \frac{C_{44}}{C} + \frac{B_{4}C_{4}}{BC} - \frac{\alpha^{2}}
{A^{2}} = - 8\pi (p + \Lambda),
\end{equation}
\begin{equation}
\label{eq8}
\frac{A_{4}B_{4}}{AB} + \frac{A_{4}C_{4}}{AC} + \frac{B_{4}C_{4}}{BC} - 
\frac{3\alpha^{2}}{A^{2}} = 8\pi (\rho - \Lambda),
\end{equation}
\begin{equation}
\label{eq9}
\frac{2A_{4}}{A} - \frac{B_{4}}{B} - \frac{C_{4}}{C} = 0.
\end{equation}
Here and in what follows the suffix $4$ by the symbols $A$, $B$, $C$ and $\rho$ denote 
differentiation with respect to $t$. The Bianchi identity ($T^{ij}_{;j} = 0$) takes the 
form
\begin{equation}
\label{eq10}
\rho_{4} + (\rho + p) \theta = 0.
\end{equation}
For complete determinacy of the system, we consider a perfect-gas equation 
of state
\begin{equation}
\label{eq11}
p = \gamma \rho, ~ ~ 0 \leq \gamma \leq 1.
\end{equation}
It is worth noting here that our approach suffers from a lack of Lagrangian 
approach. There is no known way to present a consistent Lagrangian model
satisfying the necessary conditions discussed in this paper. \\
The physical quantities expansion scalar $\theta$ and shear scalar $\sigma^{2}$
have the following expressions:
\begin{equation}
\label{eq12}
\theta = u^{i}_{;i} = \frac{A_{4}}{A} + \frac{B_{4}}{B} + \frac{C_{4}}{C}
\end{equation} 
\begin{equation}
\label{eq13}
\sigma^{2} = \frac{1}{2}\sigma_{ij}\sigma^{ij} = \frac{1}{3}\left[\theta^{2} - 
\frac{A_{4}B_{4}}{AB} - \frac{A_{4}C_{4}}{AC} - \frac{B_{4}C_{4}}{BC}\right].
\end{equation} 
Integrating Eq. (\ref{eq9}) and absorbing the integration constant into $B$ or
$C$, we obtain
\begin{equation}
\label{eq14}
A^{2} = BC
\end{equation} 
without any loss of generality. Thus, elimination of $p$ from
Eqs. (\ref{eq5}) - (\ref{eq7}) gives the condition of isotropy of pressures
\begin{equation}
\label{eq15}
2\frac{B_{44}}{B} + \left(\frac{B_{4}}{B}\right)^{2} = 2\frac{C_{44}}{C} + 
\left(\frac{C_{4}}{C}\right)^{2},
\end{equation} 
which on integration yields
\begin{equation}
\label{eq16}
\frac{B_{4}}{B} - \frac{C_{4}}{C} = \frac{k}{(BC)^{3/2}},
\end{equation}
where $k$ is a constant of integration. Hence for the metric function $B$ or 
$C$ from the above first order differential Eq. (\ref{eq16}), some scale 
transformations permit us to obtain new metric function $B$ or $C$. \\

Firstly, under the scale transformation $dt = B^{1/2}d\tau$, Eq. (\ref{eq16}) 
takes the form     
\begin{equation}
\label{eq17}
C B_{\tau} - B C_{\tau} = k C^{-1/2},
\end{equation}
where subscript represents derivative with respect to $\tau$. Considering Eq. 
(\ref{eq17}) as a linear differential equation for $B$, where $C$ is an arbitrary
function, we obtain
\begin{equation}
\label{eq18}
(i) B = k_{1} C + k C \int{\frac{d\tau}{C^{5/2}}},
\end{equation}
where $k_{1}$ is an integrating constant. Similarly, using the transformations
$dt = B^{3/2}d\tilde{\tau}$, $dt = C^{1/2}dT$, and $dt = C^{3/2} d\tilde{T}$
in Eq. (\ref{eq16}), after some algebra we obtain respectively 
\begin{equation}
\label{eq19}
(ii) B(\tilde{\tau}; k_{2}, k) = k_{2} C ~ \exp{\left(k\int{\frac{d\tilde{\tau}}
{C^{3/2}}}\right)}, 
\end{equation}
\begin{equation}
\label{eq20}
(iii) C(T; k_{3}, k) = k_{3} B - k B \int{\frac{dT}{B^{5/2}}},
\end{equation}
and
\begin{equation}
\label{eq21}
(iv) C(\tilde{T}; k_{4}, k) = k_{4} B ~ \exp{\left(k\int{\frac{d\tilde{T}}
{B^{3/2}}}\right)}, 
\end{equation}
where $k_{2}$, $k_{3}$ and $k_{4}$ are constants of integration. Thus choosing 
any given function $B$ or $C$ in cases (i), (ii), (iii) and (iv), one can obtain 
$B$ or $C$ and hence $A$ from (\ref{eq14}).

%%%%%%%%%%%%%%%%%%%%%  SECTION 3  %%%%%%%%%%%%%%%%%%%%%%%%%%%%%%%%%%%%%%%%%%
\section{Generation of new  solutions} 

We consider the following four cases:  

%%%%%%%%%%%%%%%%%%%%%%%%%%%%%%%%% SUBSECTION 3.1  %%%%%%%%%%%%%%%%%%%%%%%%%%%%%%%%%
\subsection{Case (i): Let $C = \tau^{n}$, ($n$ is a real number satisfying
$n \ne \frac{2}{5}$).}  
In this case, Eq. (\ref{eq18}) gives
\begin{equation}
\label{eq22}
B = k_{1}\tau^{n} + \frac{2k}{2 - 5n}\tau^{1 - 3n/2}
\end{equation}
and then from (\ref{eq14}), we obtain
\begin{equation}
\label{eq23}
A^{2} = k_{1}\tau^{2n} + \frac{2k}{2-5n}\tau^{1 - n/2}.
\end{equation}
Hence the metric (\ref{eq3}) reduces to the new form
\begin{equation}
\label{eq24}
ds^{2} = \left(k_{1}\tau^{n} + 2\ell \tau^{\ell_{1}}\right)[d\tau^{2} - \tau^{n} 
dx^{2}] - e^{2\alpha x}\left[\left(k_{1}\tau^{n} + 2\ell \tau^{\ell_{1}}\right)^
{2} dy^{2} + \tau^{2n} dz^{2}\right],
\end{equation}
where
\[
\ell = \frac{k}{2 - 5n} ~ and ~ \ell_{1} = 1 - \frac{3n}{2}.
\]
For this derived model (\ref{eq24}), the pressure, energy density and cosmological 
constant are given by
\[
8\pi(p + \Lambda) = \left(k_{1}\tau^{n} + 2\ell \tau^{\ell_{1}}\right)^{-3}
\biggl[-2k^{2}_{1} n(n - 1)\tau^{2n -2} - k_{1}\ell n(10 - 13n)\tau^{-(\ell_{1} + 2n)}
\] 
\begin{equation}
\label{eq25}
- \frac{\ell^{2}(4 + 4n - 11n^{2})}{2}\tau^{-3n}\biggr] + \alpha^{2}\tau^{-n}
\left(k_{1} \tau^{n} + 2\ell \tau^{\ell_{1}}\right)^{-1},
\end{equation}
\[
8\pi(\rho - \Lambda) = \left(k_{1}\tau^{n} + 2\ell \tau^{\ell_{1}}\right)^{-3}
\biggl[3k^{2}_{1} n^{2}\tau^{2n -2} + 3k_{1}\ell n(2 - n)\tau^{-(\ell_{1} + 2n)}
\] 
\begin{equation}
\label{eq26}
+ \frac{\ell^{2}(4 + 4n - 11n^{2})}{2}\tau^{-3n}\biggr] - 3\alpha^{2}\tau^{-n}
\left(k_{1} \tau^{n} + 2\ell \tau^{\ell_{1}}\right)^{-1}.
\end{equation}
Eq. (\ref{eq25}), with the use of (\ref{eq11}) and (\ref{eq26}), reduces to                    
\[
8\pi(1 + \gamma)\rho =  \left(k_{1}\tau^{n} + 2\ell \tau^{\ell_{1}}\right)^{-3}
\biggl[k^{2}_{1} n(n + 1)\tau^{2n -2} + 2k_{1}\ell n(5n - 2)\tau^{-(\ell_{1} + 2n)}\biggr]
\] 
\begin{equation}
\label{eq27}
 - 2\alpha^{2}\tau^{-n}\left(k_{1} \tau^{n} + 2\ell \tau^{\ell_{1}}\right)^{-1}
\end{equation}
Eliminating $\rho(t)$ between (\ref{eq26}) and (\ref{eq27}), we obtain 
\[
8\pi(1 + \gamma)\Lambda = \left(k_{1}\tau^{n} + 2\ell \tau^{\ell_{1}}\right)^{-3}
\biggl[-k^{2}_{1} n\{(2 + 3\gamma)n - 2\}\tau^{2n -2} - 
\]
\[
k\ell n\{10 - 13n + 3\gamma(2 - n)\}\tau^{-(\ell_{1} + 2n)} - \frac{(1 + \gamma)\ell^{2}
(4 + 4n - 11n^{2})}{2}\tau^{-3n}\biggr]
\]
\begin{equation}
\label{eq28}
 + (1 + 3\gamma)
\alpha^{2}\tau^{-n}\left(k_{1}\tau^{n} + 2\ell \tau^{\ell_{1}}\right)^{-1}.
\end{equation}

%%%%%%%%%%%%%%%%%%%%%%%%%%%%%%%%%%%%%%%%%%%%%%
\bfg[t]
%\vspace*{0.5cm}
\psfig{file=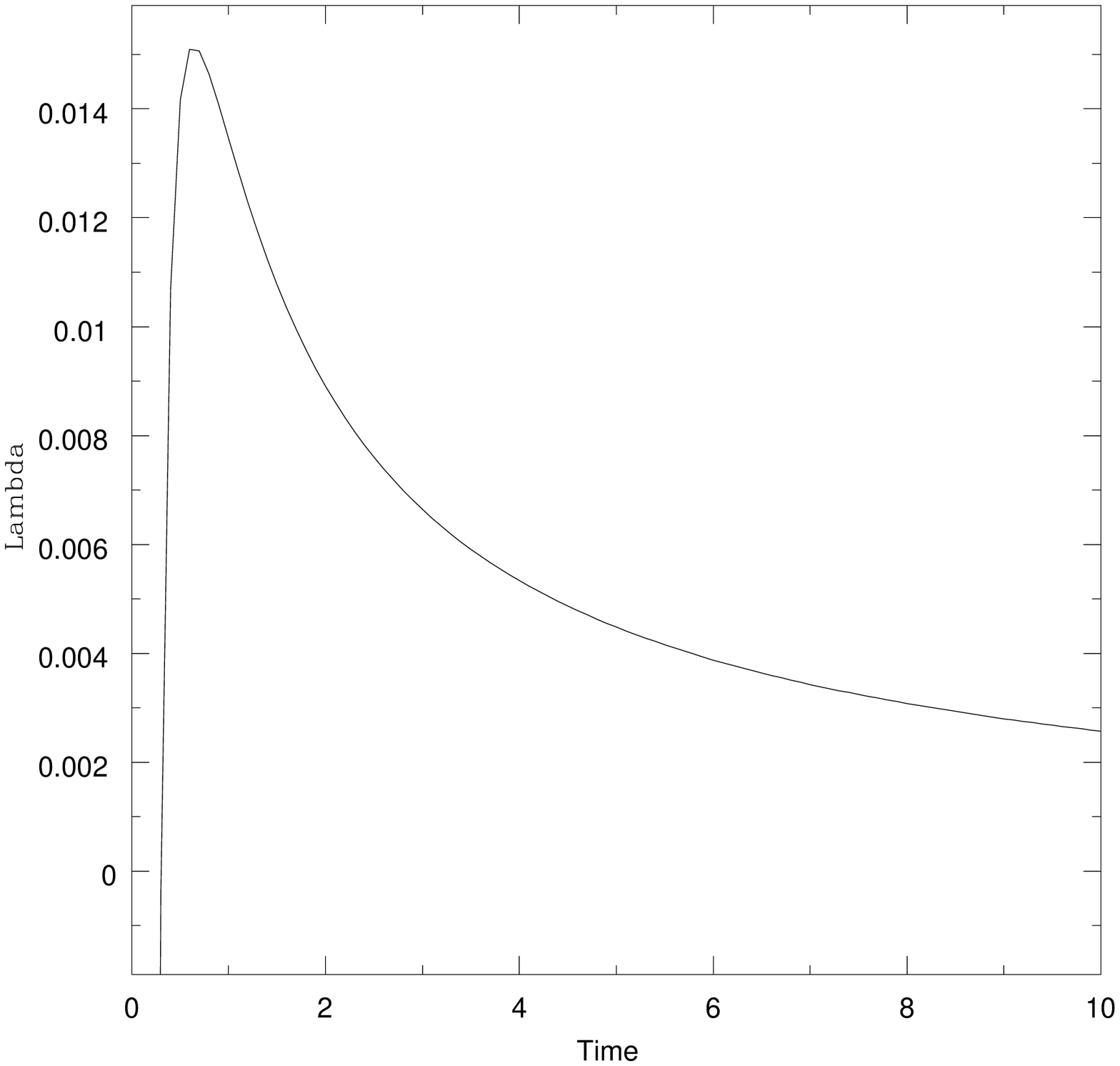,width=9cm,height=9cm}
\caption{The plot of cosmological constant $\Lambda$ versus time $\tau$ for the model 
(\ref{eq24}) in Section $3.1$ Case(i) with parameters $\alpha = 1$, $n = 0.25$, $k = 1$, 
$k_{1} = 1$, $\gamma = 0.5$.}
\efg
%\end{figure}
%%%%%%%%%%%%%%%%%%%%%%%%%%%%%%%%%%%%%%%%%%
From Eq. (\ref{eq28}), we observe that at the time of early universe the cosmological
constant ($\Lambda$) increases rapidly during a very short time and then it decreases
as time increases (see Figure 1). We also observe that the value of $\Lambda$ is 
small and positive at late times which is supported by recent type Ia supernovae 
observations [22 - 24].  \\

The metric (\ref{eq24}) is a four-parameter family of solutions to EFEs with a
perfect fluid. Using the scale transformation $dt = B^{1/2} d\tau$ in Eqs. (\ref{eq12})
and (\ref{eq13}) for this case, the scalar expansion $\theta$ and the shear $\sigma$ 
have the expressions:
\begin{equation}
\label{eq29}
\theta = 3\left[k_{1} n \tau^{n - 1} + \frac{\ell(2 - n)}{2} \tau^{-3n/2}\right]
\left(k_{1}\tau^{n} + 2\ell \tau^{\ell_{1}}\right)^{-3/2}
\end{equation} 
\begin{equation}
\label{eq30}
\sigma = \frac{1}{2} k \tau^{-3n/2}\left(k_{1}\tau^{n} + 2\ell \tau^{\ell_{1}}
\right)^{-3/2}
\end{equation}
Eqs. (\ref{eq29}) and (\ref{eq30}) lead to
\begin{equation}
\label{eq31}
\frac{\sigma}{\theta} = \frac{k}{6}\left[k_{1} n \tau^{n - \ell_{1}} + \frac{\ell(2 - n)}
{2}\right]^{-1}
\end{equation}
Now, we consider four subcases for the parameters $\Lambda$, $n$, $k$, $k_{1}$ whether 
zero or not. \\
In subcase $\Lambda = 0$, metric (\ref{eq24}) with expressions $p$, $\rho$, $\theta$ 
and $\sigma$ for this model are same as that of solution (\ref{eq18}) of Camci 
{\it et al.} [44]. \\

In subcase $\Lambda = 0$, $n = 0$, after a suitable inverse time transformation, 
we find that 
\begin{equation}
\label{eq32}
ds^{2} = dt^{2} - K(t + t_{0})^{2/3}dx^{2} - e^{2\alpha x}\left[K_{1}(t + t_{0})^
{4/3}dy^{2} + dz^{2}\right],
\end{equation}
where $t_{0}$ is a constant of integration and $K_{1} = (3k/2)^{2/3}$. The expressions 
$p$, $\rho$, $\theta$ and $\sigma$ for this model are not given here, since it is 
observed that the physical properties of this one are same as that of the solution 
(\ref{eq24}) of Ram [39]. \\

In subcase $\Lambda = 0$, $k =0$, after inverse time transformation and rescaling,
the metric (\ref{eq24}) reduces to
\begin{equation}
\label{eq33}
ds^{2} = dt^{2} - K_{2}(t + t_{1})^{\frac{4n}{n + 2}}\left[dx^{2} + e^{2\alpha x}
(dy^{2} + dz^{2})\right],
\end{equation}  
where $t_{1}$ is a constant of integration and $K_{2} = \left(\frac{n + 2}{2}\right)
^{\frac{4n}{n + 2}}$. For this solution, when $n = 1$ and $\alpha = 0$, we obtain
Einstein and de Sitter [45] dust filled universe. For $K_{2} = 1$, $t_{1} 
= 0$ and $ n = \frac{2m}{(2 - m)}$, where $m$ is a parameter in Ram's paper [39], 
the solution (\ref{eq33}) reduces to the metric (\ref{eq14}) of Ram [39]. In later case, 
if also $\alpha = 0$, then we get the Minkowski 
spacetime. \\

Now, in subcase $\Lambda = 0$, $k_{1} = 0$, after some algebra the metric (\ref{eq24})
takes the form
\begin{equation}
\label{eq34}
ds^{2} = dt^{2} - 2\ell K_{3} (t + t_{2})^{2/3}\left[dx^{2} + e^{2\alpha x} \left(
a t^{m_{1}} dy^{2} + a^{-1}t^{-m_{1}} dz^{2}\right)\right],
\end{equation} 
where $t_{2}$ is a constant, $at^{m_{1}} = 2\ell K^{\frac{2 - 5n}{2 - n}}_{3} (t + t_{2})
^{\frac{2(2 - 5n)}{3(2 - n)}}$ and $ K_{3} = \left[\frac{3(2 - n)}{4\sqrt{2 \ell}}
\right]^{2/3}$. \\
For $t_{2} = 0$, $k = \frac{2}{3}$ and $n = 0$ from (\ref{eq34}), we obtain that the 
solution (\ref{eq24}) of Ram [39]. \\ 

\noindent {\bf {Some Physical aspects of model} :} \\
The model (\ref{eq24}) has barrel singularity at $\tau = \tau_{0}$ given by
\[
\tau_{0} = \left[\frac{k_{1}(5n - 2)}{2k}\right]^{\frac{2}{(2 - 5n)}},
\]
which corresponds to $t = 0$. For $n \ne 2/5$ from (\ref{eq24}), it is observed that 
at the singularity state $\tau = \tau_{0}$, $p$, $\rho$, $\Lambda$, $\theta$ and 
$\sigma$ are infinitely large. At $t \to \infty$, which corresponds to $\tau \to 
\infty$ for $n < 2/5$ and $k> 0$, or $\tau \to 0$ for $n > 2/5$ and $k < 0$, $p$, $\rho$, 
$\Lambda$, $\theta$ and $\sigma$ vanish. Therefore, for $n \ne 2/5$, the solution
(\ref{eq24}) represents an anisotropic universe exploding from $\tau = \tau_{0}$, i.e.
$t = 0$, which expands for $0 < t < \infty$. We also find that the ratio $\sigma / \theta$
tends to a finite limit as $t \to \infty$, which means that the shear scalar does not 
tend to zero faster than the expansion. Hence the model does not approach isotropy for
large values of $t$. \\

In subcase $\Lambda = 0$, $k = 0$, the ratio (\ref{eq31}) tends to zero, then the model
approaches isotropy i.e. shear scalar $\sigma$ goes to zero. For the model (\ref{eq33}),
$p$ and $\rho$ tends to zero as $t \to \infty$; the model would give an essentially empty
universe at large time. The dominant energy condition given by Hawking and Ellis [46] 
requires that 
\begin{equation}
\label{eq35}
\rho + p \geq 0, ~ ~ \rho + 3p \geq 0
\end{equation} 
Thus, we find for the model (\ref{eq33}) that $n(2 - n) \geq 0$. Hence for the values
$0 \leq n \leq 2$, the universe (\ref{eq33}) satisfies the strong energy condition i.e.
$\rho + 3p \geq 0$. Also this model is sheer-free and expanding. \\

In subcase $\Lambda = 0$, $k_{1} = 0$, for $n \neq 2/5$, $2$, it is observed from relations
(\ref{eq25}) - (\ref{eq30}) that $p$, $\rho$, $\theta$ and $\sigma$ are infinitely large
at the singularity state $t = - t_{2}$. When $t \to \infty$, these quantities vanish.
We also find that the ratio $\sigma/ \theta$ is a constant. This shows that the cosmological
 model (\ref{eq34}) does not approach isotropy for large value of $t$. In this model the
dominant energy conditions (\ref{eq35}) are then verified for $6 - 5n - 25n^{2} \geq 0$.
Since $n \ne 2/5$, the model (\ref{eq34}) satisfies the strong energy condition for 
$-3/5 \leq n \leq 2/5$. \\
In each of subcases, all the obtained solutions (\ref{eq32}), (\ref{eq33}) and 
(\ref{eq34}) satisfy the Bianchi identity given in Eq. (\ref{eq10}).  

%%%%%%%%%%%%%%%%%%%%%%%%%%%%%%%%%  SUBSECTION 3.2 %%%%%%%%%%%%%%%%%%%%%%%%%%%%%%
\subsection{Case (ii): Let $C = \tilde{\tau}^{n}$, ($n$ is a real number 
satisfying $n \ne 2/3$).} 
In this case Eq. (\ref{eq19}) gives
\begin{equation}
\label{eq36}
B = k_{2} \tilde{\tau}^{n} \exp{\left(M\tilde{\tau}^{\ell_{1}}\right)}
\end{equation} 
and from (\ref{eq14}), we obtain
\begin{equation}
\label{eq37}
A^{2} = k_{2} \tilde{\tau}^{2n} \exp{\left(M\tilde{\tau}^{\ell_{1}}\right)}
\end{equation} 
where $M = \frac{k}{\ell_{1}}$. Hence the metric (\ref{eq3}) reduces to the form
\[
ds^{2} = \tilde{\tau}^{4(1 - \ell_{1})/3}\biggl[\tilde{\tau}^{2(1 - \ell_{1})/3} 
e^{3M\tilde{\tau}^{\ell_{1}}} d\tilde{\tau}^{2} - e^{M\tilde{\tau}^{\ell_{1}}} dx^{2}
\]  
\begin{equation}
\label{eq38}
- e^{2\alpha x}\left(e^{2M\tilde{\tau}^{\ell_{1}}} dy^{2} + dz^{2}\right)\biggr],
\end{equation}
where the constant $k_{2}$ is taken, without any loss of generality, equal to $1$.
This metric is a three-parameter family of solutions to EFEs with a perfect fluid.\\
For the above model, the distribution of matter and nonzero kinematical parameters
are obtained as  
\[
8\pi(p + \Lambda) = 2n\tilde{\tau}^{2(\ell_{1} - 2)} + 3nk\tilde{\tau}^{3\ell_{1} - 4}
\]
\begin{equation}
\label{eq39}
+ \frac{k^{2}}{2}\tilde{\tau}^{4(\ell_{1} - 1)} + \alpha^{2}\tilde{\tau}^
{4(\ell_{1} - 1)/3} e^{-3M\tilde{\tau}^{\ell_{1}}},
\end{equation} 
\[
8\pi(\rho - \Lambda) = 3n^{2}\tilde{\tau}^{2(\ell_{1} - 2)} + 3nk\tilde{\tau}^
{3\ell_{1} - 4}
\]
\begin{equation}
\label{eq40}
+ \frac{k^{2}}{2}\tilde{\tau}^{4(\ell_{1} - 1)} - 3\alpha^{2}\tilde{\tau}^
{4(\ell_{1} - 1)/3} e^{-3M\tilde{\tau}^{\ell_{1}}}.
\end{equation} 
With the use of Eq. (\ref{eq11}) in (\ref{eq39}) and (\ref{eq40}), we obtain
\[
8\pi(1 + \gamma)\rho =  n(2 + 3n)\tilde{\tau}^{2(\ell_{1} - 2)} + 6nk\tilde{\tau}^
{3\ell_{1} - 4} 
\]
\begin{equation}
\label{eq41}
+ k^{2}\tilde{\tau}^{4(\ell_{1} - 1)} - 2\alpha^{2}\tilde{\tau}^
{4(\ell_{1} - 1)/3} e^{-3M\tilde{\tau}^{\ell_{1}}}.
\end{equation}
Eliminating $\rho(t)$ between (\ref{eq40}) and (\ref{eq41}), we get
 \[
8\pi(1 + \gamma)\Lambda =  n(2 - 3n\gamma)\tilde{\tau}^{2(\ell_{1} - 2)} + 
3nk(1 - \gamma)\tilde{\tau}^{3\ell_{1} - 4} 
\]
\begin{equation}
\label{eq42}
+ \frac{k^{2}}{2}(1 - \gamma)\tilde{\tau}^{4(\ell_{1} - 1)} + \alpha^{2}(1 - 3\gamma)
\tilde{\tau}^{4(\ell_{1} - 1)/3} e^{-3M\tilde{\tau}^{\ell_{1}}}.
\end{equation} 
%%%%%%%%%%%%%%%%%%%%%%%%%%%%%%%%%%
\bfg[t]
%\vspace*{-2.5cm}
\psfig{file=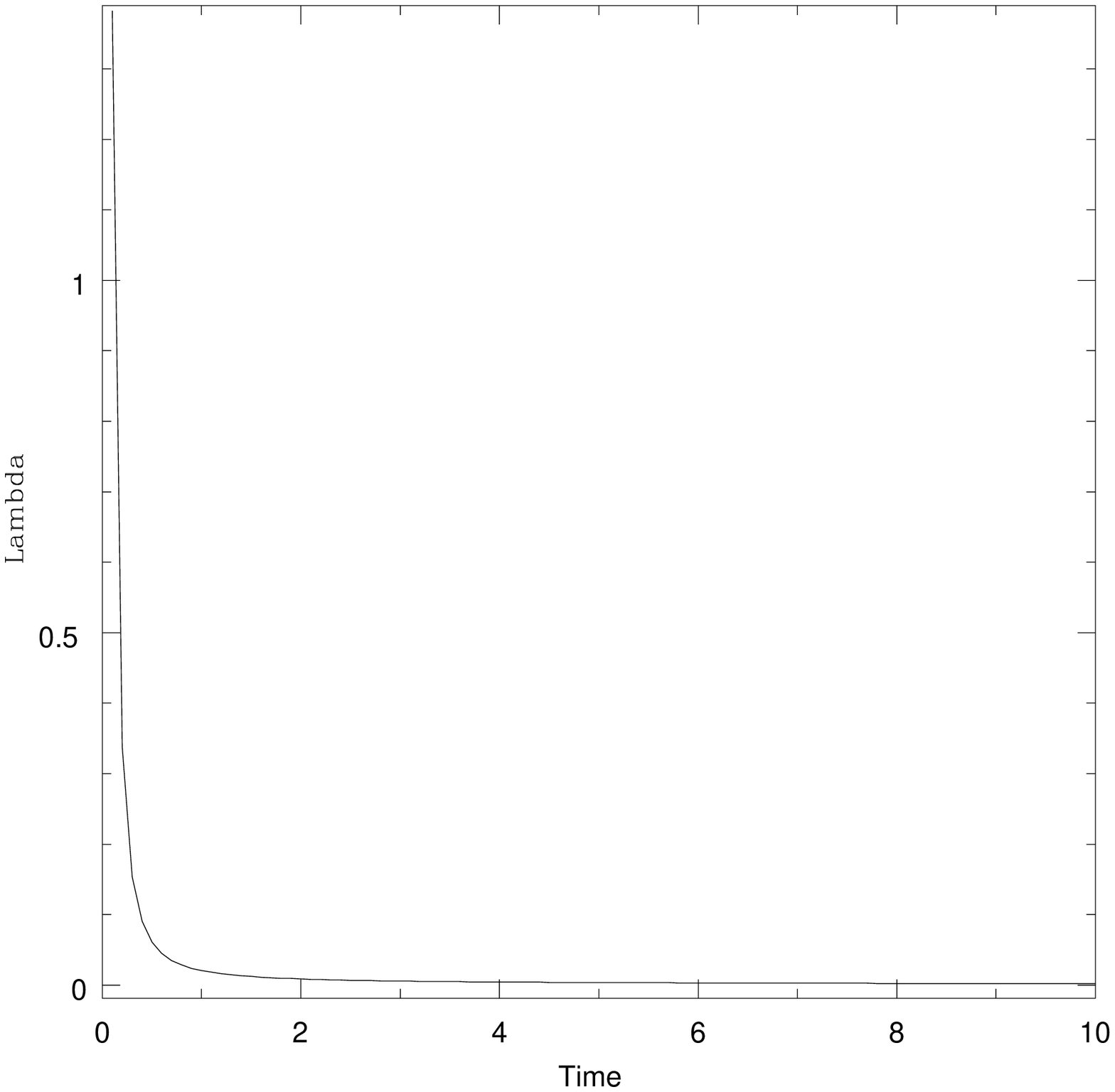,width=10cm,height=10cm}
\caption{The plot of cosmological constant $\Lambda$ versus time $\tilde{\tau}$ for 
the model (\ref{eq38}) in Section $3.2$ Case(ii) with parameters $\alpha = 1$, $n = 0.25$,
$k = 1$, $\gamma = 0.5$.}
\efg
%%%%%%%%%%%%%%%%%%%%%%%%%%%%%%%%%%%%%%%
From Eq. (\ref{eq42}), we observe that the cosmological constant ($\Lambda$) in the model
(\ref{eq38}) is a decreasing function of time (see Figure 2). We also observe that 
the value of $\Lambda$ approaches a small and positive value at late times which is 
supported by recent type Ia supernovae observations [21 - 24].  \\

The scalar of expansion $\theta$ and the shear $\sigma$ are obtained as
\begin{equation}
\label{eq43}
\theta = 3\left[n\tilde{\tau}^{\ell_{1} - 2} + \frac{k}{2}\tilde{\tau}^{2(\ell_{1} 
- 1)}\right],
\end{equation} 
\begin{equation}
\label{eq44}
\sigma = \frac{k}{2}\tilde{\tau}^{2(\ell_{1} - 1)}e^{-3M\tilde{\tau}^{\ell_{1}}},
\end{equation} 
From Eqs. (\ref{eq43}) and (\ref{eq44}), we have
\begin{equation}
\label{eq45}
\frac{\sigma}{\theta} = \frac{k}{6\left(n \tilde{\tau}^{-\ell_{1}} + \frac{k}{2}\right)}.
\end{equation} 

In subcase $\Lambda = 0$, metric (\ref{eq38}) with expressions $p$, $\rho$, $\theta$ 
and $\sigma$ for this model are same as that of solution (\ref{eq27}) of Camci 
{\it et al.} [44]. \\

In sub-case $\Lambda = 0$, $\ell_{1} = 1$ (i.e. $n = 0$), we find a similar solution to
(\ref{eq32}), and hence this subclass is omitted. For $k =0$, the ratio (\ref{eq45}) 
is zero and hence there is no anisotropy. \\
After a suitable coordinate transformation, the metric (\ref{eq38}) can be written as
\begin{equation}
\label{eq46}
ds^{2} = dt^{2} - K_{4}(t + t_{3})^{2\ell_{1}}\left[dx^{2} + e^{2\alpha x}(dy^{2} 
+ dz^{2})\right],
\end{equation} 
where $t_{3}$ is a constant and $K_{4} = \left[2/(2 - 3M_{1})\right]^{2M_{1}}$, 
$M_{1} = \frac{2n}{2 + 3n} \ne 2/3$, where $M_{1}$ is a new parameter. When $M_{1} = 0$ 
and $\ell_{1} = 0$, from (\ref{eq46}), we get the solution (\ref{eq12}) of Ram [39].\\

\noindent {\bf {Some physical aspects of model}:} \\
The model has singularity at $\tilde{\tau} \to - \infty$ for $\ell_{1} > 0$
or $\tilde{\tau} \to 0$ for $\ell_{1} < 0$, which corresponds to $t \to 0$.
It is a point type singularity for $\ell_{1} > 0$ whereas it is a cigar or a barrel 
singularity according as $\ell_{1} < 0$. At $t \to \infty$, which correspond to
$\tilde{\tau} \to \infty$ for $\ell_{1} > 0$ or $\tilde{\tau} \to 0$ for 
$\ell_{1} < 0$, from Eqs. (\ref{eq39}) - (\ref{eq45}), we obtain that for
$\ell_{1} > 0$, $p,\rho \to 0$, and $\sigma, \theta \to 0$ ($k > 0$), - $\infty$
($k < 0$); for  $\ell_{1} < 0$, similar the above ones. Then, clearly, for a
realistic universe, it must be fulfill as $\tilde{\tau} \to - \infty$, $n$ and $k$ 
are positive and $\ell_{1}$ is an odd positive number; as $\tilde{\tau} \to 0$,
$k$ is positive, and $\ell_{1}$ an even negative number. Also, since $lim_{t \to
\infty} \frac{\sigma}{\theta} \ne 0$ therefore, this model does not approach isotropy
for large values of $t$. \\

In subcase $k = 0$ for the metric (\ref{eq46}), the pressure, density and cosmological
constant are given by
\begin{equation}
\label{eq47}
8\pi (p + \Lambda) = \frac{\ell_{1}(2 - 3\ell_{1})}{(t + t_{3})^{2}} + \frac{\alpha^{2}}
{K_{4}(t + t_{3})^{2\ell_{1}}},
\end{equation} 
\begin{equation}
\label{eq48}
8\pi (\rho - \Lambda) = \frac{3\ell_{1}}{(t + t_{3})^{2}} - \frac{3\alpha^{2}}
{K_{4}(t + t_{3})^{2\ell_{1}}}.
\end{equation}
Eq. (\ref{eq47}), with the use of (\ref{eq11}) and (\ref{eq48}), reduces to 
\begin{equation}
\label{eq49}
8\pi (1 + \gamma)\rho = \frac{2\ell_{1}}{(t + t_{3})^{2}} - \frac{2\alpha^{2}}
{K_{4}(t + t_{3})^{2\ell_{1}}}.
\end{equation} 
Eliminating $\rho(t)$ between (\ref{eq48}) and (\ref{eq49}), we obtain
\begin{equation}
\label{eq50}
8\pi(1 + \gamma)\Lambda = \frac{\ell_{1}[2 - 3\ell_{1}(1 + \gamma)]}{(t + t_{3})^{2}}
+ \frac{\alpha^{2}(1 + 3\gamma)}{K_{4}(t + t_{3})^{2\ell_{1}}}.
\end{equation} 
When $\Lambda = 0$, the pressure and energy density are same as that of given in Eq.
(\ref{eq44}) of paper Camci {\it et al.} [44]. In this case, the weak and strong 
energy conditions (\ref{eq35}) for this solution are identically satisfied when 
$\ell_{1}(1 - \ell_{1}) \geq 0$ i.e. $0 \leq \ell_{1} \leq 1$. This model is shear-free 
and expanding with $\theta = \frac{3\ell_{1}}{(t + t_{3})}$.  

%%%%%%%%%%%%%%%%%%%%%%%%%%% SUBSECTION 3.3  %%%%%%%%%%%%%%%%%%%%%%%%%%%%%%%%%%
\subsection{Case (iii) : Let $B$ = $T^{n}$ ($n$ is a real number).} 
In this case Eq. (\ref{eq20}) gives
\begin{equation}
\label{eq51}
C = k_{3} T^{n} - 2\ell T^{\ell_{1}}
\end{equation}
and then from (\ref{eq14}), we obtain
\begin{equation}
\label{eq52}
A^{2} = k_{3} T^{2n} - 2\ell T^{\ell_{1} + n}
\end{equation}
Hence the metric (\ref{eq3}) takes the new form
\[
ds^{2} = \left(k_{3} T^{n} - 2\ell T^{\ell_{1}}\right)[dt^{2} - T^{n}dx^{2}] -
\]
\begin{equation}
\label{eq53}
e^{2\alpha x}\left[T^{2n}dy^{2} + \left(k_{3}T^{n} - 2\ell T^{\ell_{1}}\right)^{2}
dz^{2}\right]
\end{equation}
For four-parameters family of solution (\ref{eq53}), the physical and kinematical 
quantities are given by
\[
8\pi(p + \Lambda) = \biggl[- \frac{\ell^{2}}{2}(11n^{2} - 4n -4) T^{-3n} + \ell k_{3}
n(13n -10)T^{\ell_{1} + n} - 
\]
\begin{equation}
\label{eq54}
2k^{2}_{3} n(n - 1)T^{2n - 2}\biggr]\left(k_{3}T^{n} - 2\ell T^{\ell_{1}}\right)^{-3} 
+ \alpha^{2} T^{-n}\left(k_{3}T^{n} - 2\ell T^{\ell_{1}}\right)^{-1}, 
\end{equation}
\[
8\pi(\rho - \Lambda) = \biggl[- \frac{\ell^{2}(11n^{2} - 4n - 14)}{2} T^{-3n} - 3\ell k_{3}
n(2 - n)T^{\ell_{1} + n} + 
\]
\begin{equation}
\label{eq55}
3k^{2}_{3} n^{2} T^{2n - 2}\biggr]\left(k_{3}T^{n} - 2\ell T^{\ell_{1}}\right)^{-3} - 
3\alpha^{2} T^{-n}\left(k_{3}T^{n} - 2\ell T^{\ell_{1}}\right)^{-1}. 
\end{equation}
Eq. (\ref{eq54}), with the use of (\ref{eq11}) and (\ref{eq55}), reduces to
\[
8(1 + \gamma)\rho = \biggl[\ell^{2}(9 + 4n -11n^{2}) T^{-3} + 16\ell k_{3} n(n - 1)
T^{\ell_{1} + n} + 
\]
\begin{equation}
\label{eq56}
k^{2}_{3}n(2n + 1)T^{2n - 2}\biggr]\left(k_{3} T^{n} - 2\ell T^{\ell_{1}}\right)^{-3} 
- 2\alpha^{2}T^{-n}\left(k_{3}T^{n} - 2\ell T^{\ell_{1}}\right)^{-1}.
\end{equation}
Eliminating $\rho(t)$ between (\ref{eq55}) and (\ref{eq56}), we have
\[
8(1 + \gamma)\Lambda = \biggl[- \frac{\ell^{2}}{2}\{n(11n - 1)(1 - \gamma) - 
2(2 - 7\gamma)\}T^{-3} + 
\]
\[
\ell k_{3} n\{13n -10 + 3(2 - n)\gamma\}T^{\ell_{1} + n} - k^{2}_{3}n\{2(n - 1) - 
3n\gamma\}T^{2n - 2}\biggr] \left(k_{3} T^{n} - 2\ell T^{\ell_{1}}\right)^{-3} 
\]
\begin{equation}
\label{eq57}
+ \alpha^{2}(1 + 3\gamma)T^{-n}\left(k_{3}T^{n} - 2\ell T^{\ell_{1}}\right)^{-1}.
\end{equation}

%%%%%%%%%%%%%%%%%%%%%%%%%%%%%%%
\bfg[t]
\psfig{file=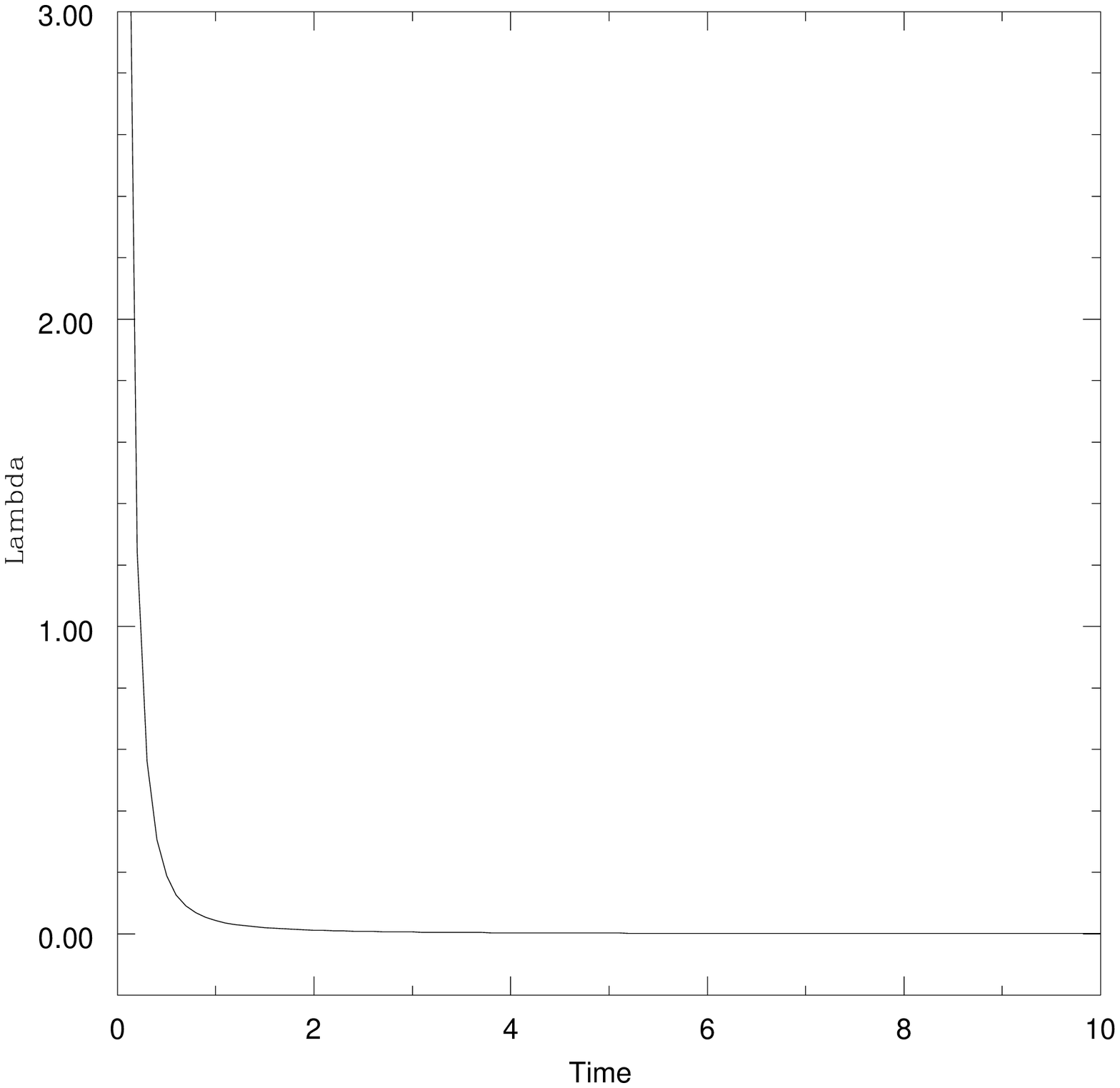,width=10cm,height=10cm}
\caption{The plot of cosmological constant $\Lambda$ versus time $T$ for the model 
(\ref{eq53}) in Section $3.3$ Case(iii) with parameters $\alpha = 1$, $n = 1$, $k = 1$, 
$k_{3} = 1.5$, $\gamma = 0.5$.}
\efg
%%%%%%%%%%%%%%%%%%%%%%%%%%%%%%%%%%%%

From Eq. (\ref{eq57}), we observe that the cosmological constant ($\Lambda$) in 
the model (\ref{eq53}) decreases as time increases. We also observe that the value 
of $\Lambda$ approaches a small and positive at late times which is supported 
by recent type Ia supernovae observations (\ref{eq21}) $-$ (\ref{eq24}). Figure $3$ 
clearly shows this behaviour of $\Lambda$ as decreasing function of time. \\

The scale of expansion and the shear are obtained as
\begin{equation}
\label{eq58}
\theta = 3\left[\frac{\ell(n - 2)}{2} T^{-3n/2} + k_{3} n T^{n - 1}\right]
\left(k_{3} T^{n} - 2\ell T^{\ell_{1}}\right)^{-3/2},
\end{equation}
\begin{equation}
\label{eq59}
\sigma = \frac{kT^{-3n/2}}{2}\left(k_{3} T^{n} - 2\ell T^{\ell_{1}}\right)^{-3/2}.
\end{equation}
From (\ref{eq58}) and (\ref{eq59}), we get
\begin{equation}
\label{eq60}
\frac{\sigma}{\theta} = \frac{k}{6}\left[k_{3} nT^{-\ell_{1} + n} + \frac{\ell (n - 2)}
{2}\right]^{-1}.
\end{equation}
In subcase $\Lambda = 0$, metric (\ref{eq53}) with expressions $p$, $\rho$, $\theta$ 
and $\sigma$ for this model are same as that of solution (\ref{eq34}) of 
Camci {\it et al.} [44]. \\

In subcase $\Lambda = 0$, $n = 0$, after an inverse transformation, metric (\ref{eq53})
reduces to the form
\begin{equation}
\label{eq61}
ds^{2} = dt^{2} - K_{5}(t + t_{4})^{2/3}dx^{2} - e^{2\alpha x}\left[dy^{2} + K^{2}_{5}
(t + t_{4})^{4/3}dz^{2}\right],
\end{equation}
where $t_{4}$ is an integrating constant. This model is different from the model 
(\ref{eq32}) by a change of scale. \\

In subcase $\Lambda = 0$, $k = 0$, same model as (\ref{eq33}) is obtained. \\

Further in subcase  $\Lambda = 0$, $k_{3} = 0$, we see that the metric (\ref{eq53})
takes the form
\begin{equation}
\label{eq62}
ds^{2} = dt^{2} - 2\ell K_{6}(t + t_{5})^{2/3}\left[dx^{2} + e^{2\alpha x}
\left(b t^{m_{2}}dy^{2} + b^{-1}t^{-m_{2}} dz^{2}\right)\right],
\end{equation}
where $t_{5}$ is a constant, $bt^{m_{2}} = 2\ell K^{\frac{2 - 5n}{2 - n}}_{6}
(t + t_{5})^{\frac{2(2 - 5n)}{3(2 - n)}}$ and $K_{6} = \left[\frac{3(2 - n)}
{4\sqrt{2 \ell}}\right]^{2/3}$. This metric is only different from (\ref{eq33}) by
a change of sign. Also, in each of subcase the physical and kinematical properties
of obtained metric are same as that of case(i). Therefore, we do not consider here them. \\

%%%%%%%%%%%%%%%%%%%%%%%%%%% SUBSECTION 3.4  %%%%%%%%%%%%%%%%%%%%%%%%%%%%%%%%%%
\subsection{Case (iv) : Let $B$ = $\tilde{\tau}^{n}$, where $n$ is any real number.} 
In this case Eq. (\ref{eq21}) gives
\begin{equation}
\label{eq63}
C = k_{4} \tilde{\tau}^{n}\exp{\left(\frac{k}{\ell_{1}} \tilde{\tau}^{\ell_{1}}
\right)}
\end{equation}
and then from (\ref{eq14}), we obtain
\begin{equation}
\label{eq64}
A^{2} = k_{4}\tilde{\tau}^{2n}\exp{\left(\frac{k}{\ell_{1}} \tilde{\tau}^{\ell_{1}}
\right)} 
\end{equation}
Hence the metric (\ref{eq3}) reduces to
\[
ds^{2} = \tilde{\tau}^{2n}\exp{\left(\frac{k}{\ell_{1}} \tilde{\tau}^{\ell_{1}}\right)}
\left[\tilde{\tau}^{n}\exp{\left(\frac{2k}{\ell_{1}} \tilde{\tau}^{\ell_{1}}\right)}
- dx^{2}\right]
\]
\begin{equation}
\label{eq65}
- e^{2 \alpha x}\left[dy^{2} + \exp{\left(\frac{2k}{\ell_{1}} \tilde{\tau}^{\ell_{1}}
\right)} - dz^{2}\right],
\end{equation}
where, without any loss of generality, the constant $k_{4}$ is taken equal to $1$. 
Expressions for physical and kinematical parameters for the model (\ref{eq65}) are 
not given here, but it is observed that the properties of the metric (\ref{eq65}) 
are same as that of the solution (\ref{eq38}), i.e. the case (ii). \\

%%%%%%%%%%%%%%%%%%%%%  SECTION 4  %%%%%%%%%%%%%%%%%%%%%%%%%%%%%%%%%%%%%%%%%%
\section{Concluding remarks} 
In this paper we have described a new exact solutions of EFES for Bianchi type V 
spacetime with a perfect fluid as the source of matter and cosmological term $\Lambda$ 
varying with time. Using a generation technique followed by Camci {\it et al.} [44],
it is shown that the Einstein's field equations are solvable for any arbitrary cosmic 
scale function. Starting from particular cosmic functions, new classes of spatially
homogeneous and anisotropic cosmological models have been investigated for which the
fluids are acceleration and rotation free but they do have expansion and shear. For
$\alpha = 0$ in the metric (\ref{eq1}), we obtained metrics as LRS Bianchi type I model
(Hajj-Boutros [37], Hajj-Boutros and Sfeila [38], Ram [39], 
Mazumder [40] and Pradhan and Kumar [41]). It is also seen that the 
solutions obtained by Camci {\it et al.} [44] and  Ram [39] are particular
of our solutions. \\
The cosmological constants in all models given in Section $3$ are decreasing functions
of time and they all approach a small positive value at late times which are supported 
by the results from recent supernova observations recently obtained by the High-z Supernova 
Team and Supernova Cosmological Project (Garnavich {\it et al.} [21], Perlmutter 
{\it et al.} [22], Riess {\it et al.} [23], Schmidt {\it et al.} [24]).

\bigskip
{\small One of the authors (A. Pradhan) would like to thank the Inter-University
Centre for Astronomy and Astrophysics, Pune, India for providing facility under Associateship 
Programme where part of this work was carried out. Authors are also thankful to C. S. Stalin 
for his help in plotting the figures.}
                 %for \end{thebibliography}
\end{document}